\documentclass[conference]{IEEEtran}
\usepackage{multirow}
\ifCLASSINFOpdf
   \usepackage[pdftex]{graphicx}
   
   \graphicspath{{./pdf/}{./jpeg/}}
   \DeclareGraphicsExtensions{.pdf,.jpeg,.png}
\else
   \usepackage[dvips]{graphicx}
   \graphicspath{{./eps/}}
   \DeclareGraphicsExtensions{.eps}
\fi
\usepackage{booktabs}
\usepackage{mathtools}
\usepackage{color}
\usepackage{algorithm}
\usepackage{algpseudocode}

\usepackage{times}
\usepackage{epsfig}
\usepackage{graphicx}
\usepackage{amsmath}
\usepackage{amssymb}
\usepackage{romannum}
\usepackage{mathrsfs}
\usepackage{color}
\usepackage{caption}
\usepackage{subcaption}

\usepackage{hyperref}
\hypersetup{
    colorlinks=true,
    linkcolor=blue,
    urlcolor=blue,
    citecolor=blue
}
\usepackage{multirow}
\usepackage{algorithm,algpseudocode}

\usepackage{url}

\begin{document}
%
\title{Enhancing Learning Path Recommendation via Multi-task Learning }


\author{\IEEEauthorblockN{Afsana Nasrin, Lijun Qian, Pamela Obiomon and Xishuang Dong}
\IEEEauthorblockA{Department of Electrical and Computer Engineering \\ 
Roy G. Perry College of Engineering\\ 
Prairie View A\&M University \\
Prairie View, TX 77446, USA \\
Email: anasrin@pvamu.edu, liqian@pvamu.edu, phobiomon@pvamu.edu, xidong@pvamu.edu}

}


\maketitle

\begin{abstract}

Personalized learning is a student-centered educational approach that adapts content, pace, and assessment to meet each learner’s unique needs. As the key technique to implement the personalized learning, learning path recommendation sequentially recommends personalized learning items such as lectures and exercises. Advances in deep learning, particularly deep reinforcement learning, have made modeling such recommendations more practical and effective. This paper proposes a multi-task LSTM model that enhances learning path recommendation by leveraging shared information across tasks. The approach reframes learning path recommendation as a sequence-to-sequence (Seq2Seq) prediction problem, generating personalized learning paths from a learner’s historical interactions. The model uses a shared LSTM layer to capture common features for both learning path recommendation and deep knowledge tracing, along with task-specific LSTM layers for each objective. To avoid redundant recommendations, a non-repeat loss penalizes repeated items within the recommended learning path. Experiments on the ASSIST09 dataset show that the proposed model significantly outperforms baseline methods for the learning path recommendation.

\end{abstract}

\begin{IEEEkeywords} Learning Path Recommendation, Multi Task Learning; \end {IEEEkeywords}

%
\IEEEpeerreviewmaketitle

\section{Introduction}
\label{sec1}

Personalized learning is a student-centered approach in education that recognizes and accommodates the unique attributes of each learner by tailoring educational content, speed, and assessment to fulfill individual needs~\cite{ayeni2024ai}.  Fundamental components include learner profiles, cognitive styles, learning objectives, adjustable pacing, self-regulation, and technology that enables customization, such as intelligent learning environments, intelligent tutoring systems, and data mining~\cite{shemshack2021comprehensive, yuyun2023components, wibawa2025personalized}. The applications of personalized learning include not just conventional education but also AI assisted health coaching for chronic diseases, adaptive mathematics platforms using gaming, personalized health and wellness recommendation and customized dietary advice during the menstrual cycle~\cite{sqalli2019ai}~\cite{bang2023efficacy}~\cite{sathya2024fitness}~\cite{logapriya2024optimized}. To implement the personalization learning, learning path recommendation has emerged as a central strategy for guiding learners through tailored sequences of content.

Learning path recommendation can sequentially recommend personalized learning items (e.g., lectures, exercises) to address the unique needs of each student~\cite{mansouri2023full, zhu2021survey}. For instance, given a learner’s historical learning items (or records), the goal is to recommend a learning path with $N$ items that maximizes the overall learning effectiveness. Unlike a static curriculum, learning path recommendation provides a tailored sequence that not only reduces the cost of learning but also helps learners master complex concepts more effectively. This area has seen significant technological progress over the past decade. Traditional approaches such as rule-based heuristics~\cite{khalid2022literature}, collaborative filtering, and knowledge graphs~\cite{meng2021ld} provided encouraging starting points but are insufficient for scaling up and adapting to complex learner behaviors. 

Recent advances in deep learning, specifically in deep reinforcement learning~\cite{li2017deep}, have made modeling learning path recommendation more practical and expressive. Chen \textit{et al.}  proposed a novel framework named Set-to-Sequence Ranking-based Concept-aware Learning Path Recommendation (SRC)~\cite{chen2023set}, which formulates the recommendation task under a set-to-sequence paradigm by using a concept-aware encoder to capture correlations among input learning concepts and a decoder with an attention mechanism to sequentially generate an optimized learning path. Zhang \textit{et al.} proposed a novel method named Difficulty-constrained Learning Path Recommendation (DLPR)~\cite{zhang2024item}, which is aware of item difficulty by explicitly categorizing items into learning and practice types, constructing a hierarchical graph to model and leverage item difficulty, and designing a Difficulty-driven Hierarchical Reinforcement Learning (DHRL) framework to generate learning paths that balance efficiency and smoothness. Zhang \textit{et al.} introduced the privileged feature distillation technique to build Privileged Knowledge State Distillation (PKSD) framework~\cite{li2024privileged}, which enables the reinforcement learning agent to leverage the actual knowledge state as privileged information in the state encoding to better tailor recommendations to individual needs.

This paper proposes a multi-task LSTM model to enhance learning path recommendation by fully leveraging shared information across different tasks. It begins by redefining learning path recommendation as a sequence-to-sequence (Seq2Seq) prediction problem that generates a learning path based on a learner’s historical sequence. The model employs multi-task deep learning~\cite{ruder2017overview} to jointly address two tasks: learning path recommendation and deep knowledge tracing~\cite{piech2015deep}. Its architecture includes a shared LSTM layer to extract common features for both tasks, along with two task-specific LSTM layers to individually learn the objectives of learning path recommendation and deep knowledge tracing. To prevent redundant recommendations, a non-repeat loss is introduced to penalize repetitions within the generated learning path. Extensive experiments on the ASSIST09 dataset demonstrate that the proposed model significantly improves the performance of learning path recommendation.

Our main contribution are summarized as follows:
\begin{enumerate}
  \item We propose a multi-task learning path recommendation model that not only anticipates the next concept in a learning sequence but also predicts how likely the suggested path is to improve learner performance through deep knowledge tracing. Moreover, the model employs a shared LSTM encoder to learn both tasks simultaneously, which reduces task-specific overfitting and improves generalization.

 \item The proposed model is comprehensively validated on the ASSIST09 dataset. Specifically, it thoroughly examines how the length of learning paths affects recommendation performance by comparing it with six baseline models based on recurrent neural networks (RNNs)~\cite{salehinejad2017recent} and attention mechanisms~\cite{de2022attention}. Experimental results indicate that the proposed model consistently outperforms the baselines across various evaluation metrics, such as AUC and F1 scores.
 
\end{enumerate}

\section{Task}
\label{sec2}

Learning path recommendation is the key technique to implement personalized learning. Generally, it can be defined as a reinforcement learning process that is to maximizes the student's expected learning performance. For instance, given a student's historical learning sequence $X$ and a set of target concepts $T$, the task is to select $m$ distinct concepts from $T$ and generate a learning path $\pi$ that maximizes the student's expected learning performance upon completing the path. However, in this study, we redefined it as a sequence to sequence (Seq2Seq) prediction problem, where the input $X = \{x_1, x_2, ... , x_n\}$ is a user history learning sequence and the output $Y = \{y_1, y_2, ... , y_t, ... , y_m\}$ is the recommended learning path $\pi$. Here, $n$ and $m$ denote the lengths of input and output sequences, respectively.

The task aims to learning a Seq2Seq model to conduct $Y$ conditioned on $X$ as below.

\begin{equation}
P_\theta(Y|X) = \prod_{t=1}^m  P(y_t|y_{< t}, X).
\end{equation}

where each $y_t$ is drawn from set of target concepts $T$, and $\theta$ refers to the model parameters.

\section{Methodology}
\label{sec3}

This paper applied multi-task deep learning~\cite{ruder2017overview} to enhance learning path recommendation.

\subsection{Multi-Task Deep Learning}

Multi-Task Learning (MTL) is a machine learning paradigm that aims to leverage useful information contained in multiple related tasks to improve the generalization performance of all tasks~\cite{zhang2021survey}. Let there be a task set $\mathcal{T} = \{ \tau_1 , \tau_2, \ldots , \tau_t, \ldots , \tau_n \}$ with $n$ tasks. For each $\tau_t$, the input space, output space, and dataset are denoted by $\mathcal{X}$, $\mathcal{Y}_t$, and $D_t = {(x_i^t, y_i^t)}_{i = 1}^{N_t}$, respectively. Specifically, all tasks share the same input space $\mathcal{X}$. A typical MTL model includes a shared feature extractor $\mathcal{H} = f_\theta(\mathcal{X})$ and a task-specific model $\mathcal{Y}_t = g_{\phi_{t}}(\mathcal{H})$, where $\theta$ and $\phi_t$ are the parameters learned for the feature extractor and the task-specific model, respectively. Thus, the prediction for task $\tau_t$ is $\hat{y}^t = g_{\phi_{t}}(f_\theta(\mathcal{X}))$. The general MTL objective is to jointly minimize the sum of task losses:

\begin{equation}
\min_{\theta, \{\phi_t\}_{t=1}^T} \sum_{t=1}^T \lambda_t \mathcal{L}_t \left( \{ y_i^t, \hat{y}_i^t \}_{i=1}^{N_t} \right)
\quad \text{where} \quad
\hat{y}_i^t = g_{\phi_t} (f_\theta (x_i^t)).
\end{equation}

where $\mathcal{L}_t$ is the loss function for task $\tau_t$ (e.g., cross-entropy, MSE) while $\lambda_t $ is a weighting factor that controls the relative importance of each task. It has been applied across a wide range of machine learning domains, from speech recognition~\cite{deng2013new} and drug discovery~\cite{ramsundar2015massively} to computer vision~\cite{girshick2015fast} and natural language processing~\cite{collobert2008unified}.

Multi-Task Deep Learning (MTDL) implements MTL using deep neural networks~\cite{ruder2017overview} to automatically learn shared, hierarchical representations directly from raw data, whereas standard MTL typically relies on manual feature engineering or shallow models with explicit parameter sharing. In MTDL, tasks share a deep feature extractor while having task-specific output layers, enabling end-to-end training on complex, high-dimensional inputs such as images, text, or sequences.

\subsection{Proposed Method}

This study implements multi-task deep learning based on Long Short-Term Memory (LSTM) neural networks to enhance learning path recommendation. Two tasks are involved: learning path recommendation and deep knowledge tracing~\cite{piech2015deep}. The learning path recommendation task is defined as a Seq2Seq prediction problem, as described in the previous section, while deep knowledge tracing is formulated as a binary classification problem that predicts whether the learning outcome will be a success or a failure.

The architecture of the proposal method is composed of a shared LSTM layer and two task-specific LSTM layers for learning path recommendation and deep knowledge tracing, respectively.  Both tasks share an LSTM layer that maps an input sequence to hidden representations:

\begin{equation}
H_i^t = LSTM_\theta(X_i^t) = [h_{i1}^t, ... , h_{iL}^t]
\end{equation}

where $\theta$ are the shared LSTM parameters and $H_i^t$ is the sequence of hidden states. $L$ is the number of hidden LSTM unit. Afterwards, each task has its own LSTM layer to generate the output:

\begin{equation}
\hat{Y}_i^t = g_{\phi_t}(H_i^t)
\end{equation}

where $\phi_t$ are the parameters of the task-specific layer. Following, the loss of these two tasks are defined as following.

1. Loss for learning path recommendation: this task is defined as the Seq2Seq problem that is to predict a learning path from a set of target concepts. In other words, it can be viewed as a multi-class classification problem that predicts the next $k$ target concepts from $T$. At each prediction step $i \in ({1, \ldots\, k}) $, the model selects a single concept based on the probability distribution $\hat{p}^{(i)} \in \mathbb{R}^{|T|}$. The true class at step $t$ is $y^{(i)} \in \{0, 1, \ldots, |T|-1\}$. Then, the loss is:

\begin{equation}
\mathcal{L}_{\text{CE}} = \frac{1}{k} \sum_{i=1}^{k} -\log\left( \hat{p}^{(i)}_{y^{(i)}} \right)
\end{equation}

where $\hat{p}^{(i)} \in \mathbb{R}^{|T|}$ denotes the predicted probability at step $t$.

2. Loss for deep knowledge tracing: it is formulated as a binary classification problem that predicts the likelihood of the model to correctly answers all recommended problems in the learning path.  Let $\hat{y}^{(i)} \in [0, 1]$ be the predicted probability at step $i$ and $y^{(i)} \in \{0, 1\}$ be the ground truth. The binary cross entropy loss computed as:

\begin{equation}
\mathcal{L}_{\text{BCE}} = 
\frac{1}{k} \sum_{i=1}^{k} \left[ -y^{(i)} \log(\hat{y}^{(i)}) - (1 - y^{(i)}) \log(1 - \hat{y}^{(i)}) \right]
\end{equation}

Moreover, we introduced non-repeat loss term for the learning path prediction and performance prediction so that no concept can appear multiple times in the recommended path. The constraint is motivated by the principle to prioritize diversity and convergence. Let $\hat{Y} = [\hat{y}_1, \hat{y}_2, \dots, \hat{y}_L] $ denote the total number of predicted concepts of length $L$. $\text{unique}(\hat{Y})$ denotes distinct concepts in $\hat{Y}$.  The repetition loss, $L_{rep}$ can be defined as:
 
 \begin{equation}
     \mathcal{L}_{\text{rep}}(\hat{Y}) = L - |\text{unique}(\hat{Y})|
 \end{equation}

This loss term enforces a penalty proportionate with the frequency of repeated concepts, leading the model to generate paths that have better conceptual diversity. 
 
 To this end, the overall training goal combines this penalty with the task-specific losses for path prediction and performance assessment.  The total loss is expressed as:
 
 \begin{equation}
\mathcal{L}_{\text{total}} = \mathcal{L}_{\text{CE}} + \lambda_1 \cdot \mathcal{L}_{\text{BCE}} + \lambda_2 \cdot \mathcal{L}_{\text{rep}}
 \end{equation}

where,
$\mathcal{L}_{\text{CE}}$ is the loss for learning path recommendation,  
$\mathcal{L}_{\text{BCE}}$ is the binary cross-entropy loss for deep knowledge tracing,  
$\lambda_1$ controls the contribution of the performance of deep knowledge tracing,  
$\lambda_2$ governs the strength of the no-repeat penalty.

\section{Experiments}
\label{sec4}

\begin{table}[htp!]
\centering
\caption{Experiment setup for the learning path recommendation.}
\label{tab:hyperparameters}
\begin{tabular}{lc}

\toprule
\textbf{Parameter} & \textbf{Value} \\ \midrule
Epochs                          & 100                    \\
Optimizer                       & Adam                   \\
Batch size                      & 32                     \\
Learning rate                   & $10^{\text{-3}}$                  \\
Embedding dimensions             & 128                    \\
Number of hidden units                    & 64                     \\
Length of history learning sequences         & 10                     \\
Length of learning paths               & 3, 5, 7, 9             \\
\bottomrule
\end{tabular}
\end{table}

\subsection{Dataset}

We use the 2009–2010 ASSISTments Skill Builder dataset (Assist09)~\cite{feng2009addressing}\footnote{\url{https://sites.google.com/site/assistmentsdata/home/2009-2010-assistment-data}}, a widely used open-source benchmark in educational data mining. This dataset consists of detailed student interaction logs collected from an online tutoring system. Each interaction records a student’s response to a specific problem, along with metadata such as correctness, response latency, and problem-to-skill mappings. Each log entry includes fields such as order ID, problem ID, correctness, attempt count, and milliseconds to first response. These attributes capture the reference to the original problem, whether the student answered correctly on the first attempt, the number of attempts, and the time taken to respond. All interactions are grouped by user\_id and sorted chronologically using order ID. For each student, we extract a sequence of attempted problems,  represented as $X = \{p_1, p_2, p_3, ... , p_i, ... , p_n\}$, where $p_i$ denotes the problem ID attempted at time step $i$. The ground truth for the learning path recommendation task is   $Y = \{p_{n+1}, p_{n+2}, p_{n+3}, ... , p_{n+L}\}$, where $L$ indicates the length of the recommended learning path.

\begin{table*}[ht]
\centering
\caption{Performance comparison on the learning path recommendation through various evaluation metrics.}
\label{tab:combined-performance}
\begin{tabular}{lcccc} \hline
 \textbf{Model} & \textbf{Accuracy} & \textbf{F1 Score} & \textbf{Precision} & \textbf{Recall} \\
\hline
Standard RNN & 0.2401 & 0.2197 & 0.2313 & 0.2401 \\
 LSTM & 0.2891 & 0.2714 & 0.2928 & 0.2891 \\ \hline
Seq2Seq + Standard RNN & 0.2081 & 0.1873 & 0.1994 & 0.2081 \\ 
Seq2Seq + LSTM & 0.3064 & 0.2934 & 0.3112 & 0.3064 \\ \hline
Seq2Seq + Standard RNN + Attention & 0.2602 & 0.2444 & 0.2657 & 0.2602\\
Seq2Seq + LSTM + Attention & 0.3060 & 0.2932 & 0.3104 & 0.3060 \\ \hline
\textbf{Proposed method} & \textbf{0.3489} & \textbf{0.3241} & \textbf{0.3202} & \textbf{0.3489} \\
\hline
\end{tabular}
\end{table*}

\subsection{Experiment Setup}

Table~\ref{tab:hyperparameters} outlines the experimental setup used for the learning path recommendation task. The training configuration includes 100 epochs with the Adam optimizer, a batch size of 32, and a learning rate set to $10^{-3}$ . The model uses embedding dimensions of size 128 and 64 hidden units to capture sequential patterns in history learning sequences. Each student’s learning history is represented by a sequence length of 10, while the lengths of the recommended learning paths evaluated in the experiments are set to 3, 5, 7, and 9. These settings define the training and evaluation conditions for comparing the performance between the baselines and the proposed method.

Baseline models include:

\begin{itemize}
\item \textbf{Standard RNN} refers to a basic recurrent neural network that processes sequences by updating a hidden state step by step to capture dependencies between elements. 
\item \textbf{LSTM} is an advanced type of recurrent neural network designed to handle long-range dependencies in sequential data more effectively than a standard RNN. It uses special gating mechanisms—input, output, and forget gates—to control the flow of information, allowing it to retain or discard information over long sequences and mitigate the vanishing gradient problem.
\item \textbf{Seq2Seq + Standard RNN} combines the Seq2Seq framework with a standard RNN as its encoder and decoder. In this setup, an RNN first encodes an input sequence into a fixed-length hidden representation, which another RNN then decodes step by step to generate an output sequence. 
\item \textbf{Seq2Seq + LSTM} uses the Seq2Seq framework with LSTM networks as both the encoder and the decoder. The encoder LSTM reads and compresses the input sequence into a context vector, and the decoder LSTM generates the output sequence step by step from this context. 
\item \textbf{Seq2Seq + Standard RNN + Attention} extends the basic Seq2Seq framework by adding an attention mechanism to a standard RNN encoder-decoder architecture. In this setup, the standard RNN encodes the input sequence, but instead of relying solely on a single fixed-length context vector, the attention mechanism allows the decoder RNN to dynamically focus on different parts of the input sequence at each output step. 
\item \textbf{Seq2Seq + LSTM + Attention} combines the strengths of LSTM cells and an attention mechanism within the  Seq2Seq framework. This design helps the model handle variable-length sequences and capture long-range dependencies more effectively than using a plain Seq2Seq or standard RNN alone, making it well-suited for complex tasks like translation, summarization, or learning path recommendation.
\end{itemize}

\subsection{Evaluation Metrics}
This study evaluates the effectiveness of various learning path recommendation models in predicting the learning path by leveraging their historical practices as predictive features.  Performance evaluation is conducted using accuracy, precision, recall, and the Area Under the Receiver Operating Characteristic Curve (AUC).

\begin{equation} 
Accuracy = \frac{TP+TN}{TP+FP+TN+FN} 
\end{equation}

Accuracy, a widely used metric in classification tasks, quantifies the proportion of correctly classified instances, including both true positives and true negatives, relative to the total number of instances evaluated. It provides a straightforward assessment of overall prediction performance.

\begin{equation} 
\text{Precision}  = \frac{TP}{TP+FP} 
\end{equation} 

\begin{equation} 
\text{Recall} = \frac{TP}{TP+FN} 
\end{equation}

\begin{equation} 
\text{F1~Score} = \frac{2 \times \text{Precision} \times \text{Recall}}{\text{Precision} + \text{Recall}}
\end{equation}

where True Positive (TP), False Positive (FP), True Negative (TN), and False Negative (FN) are derived from the confusion matrix:

\begin{itemize} 
\item TP: Correctly predicted a learning item for the learning path successfully. 
\item TN: Correctly predicted a learning item that does not belong to the learning path. 
\item FP: Incorrectly predicted a learning item for the learning path successfully. 
\item FN: Incorrectly predicted a learning item that does not belong to the learning path. \end{itemize}

Precision measures the proportion of correctly predicted learning items out of all learning items predicted to the learning path. Recall measures the proportion of correctly predicted learning items  out of all actual learning items in the learning path. 

\begin{figure*}[htbp]
  \centering
  \includegraphics[width=0.97\textwidth]{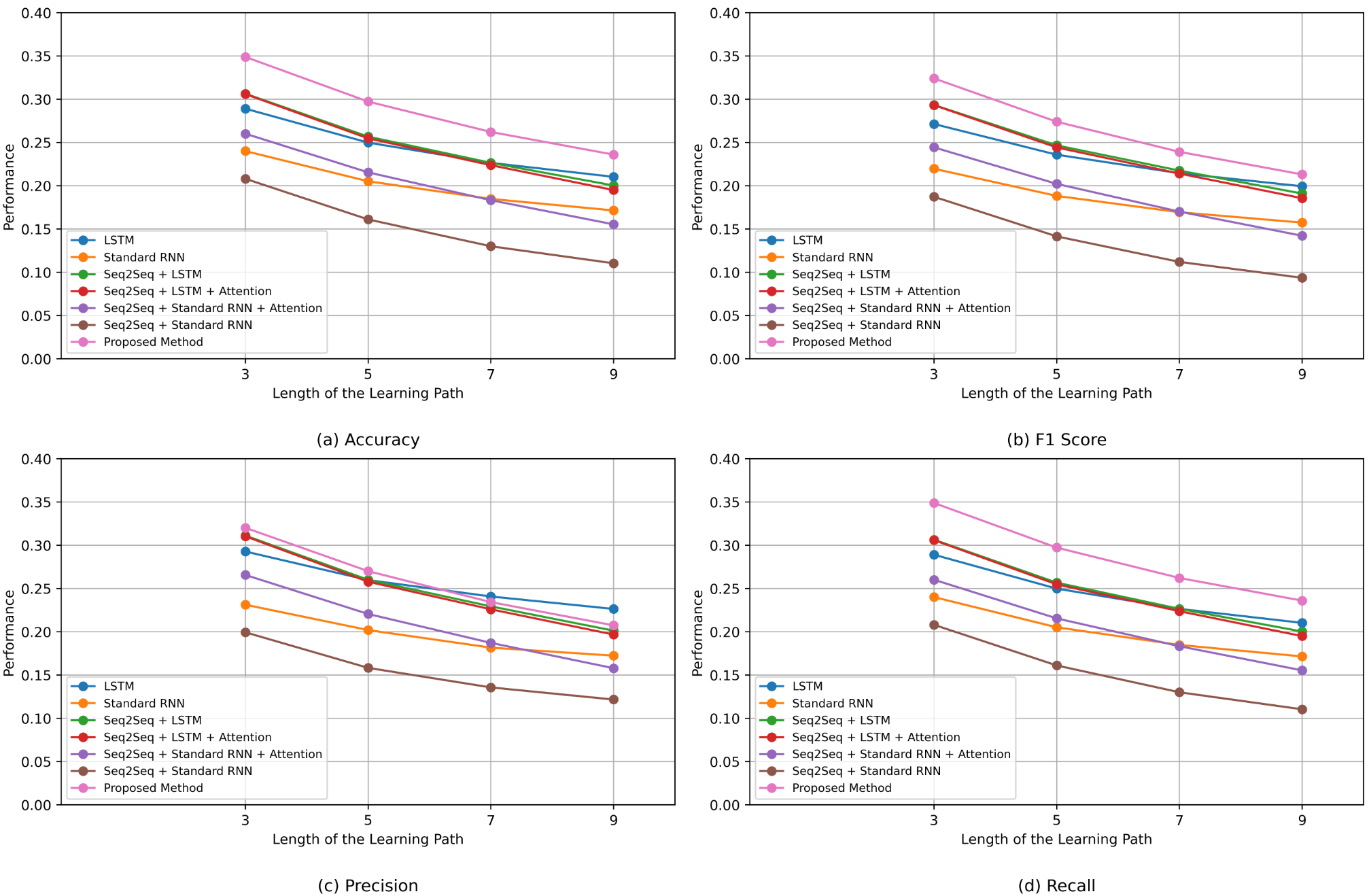}
  \caption{Performance comparison on impact of various learning path lengths.}
  \label{fig:evaluation-metrics}
\end{figure*}

\subsection{Results and Discussions}

\subsubsection{Performance comparison between baselines and the proposed method}

Table~\ref{tab:combined-performance} highlights the superiority of the proposed method over all baseline models for the learning path recommendation task. The proposed method achieves the highest performance across all evaluation metrics: an Accuracy of 0.3489, F1 Score of 0.3241, Precision of 0.3202, and Recall of 0.3489. This consistent improvement indicates that the proposed approach introduces significant enhancements that enable more effective modeling of student learning paths compared to traditional RNN, LSTM, and even Seq2Seq-based models.

When examining the progression from simpler models to more advanced configurations, it is observed that adding Seq2Seq and attention mechanisms incrementally improves performance, particularly when paired with LSTM rather than a standard RNN. For example, moving from a basic LSTM to Seq2Seq + LSTM yields better performances, and integrating attention provides further, though modest, gains. However, none of these configurations match the proposed method, which suggests that it likely incorporates additional architectural innovations or optimizations beyond simply layering Seq2Seq and attention on top of LSTM.

The experimental results underscore the valuable contributions of the multi-task LSTM aspect within the proposed method. Compared to single-task baselines like the standard LSTM or Seq2Seq + LSTM, the proposed method achieves substantial gains in all metrics, demonstrating that jointly optimizing multiple related tasks helps the model capture richer patterns in student learning behaviors. This multi-task approach likely enables the LSTM to share representations between the learning path recommendation and the deep knowledge tracing, leading to better generalization and more robust recommendations. Overall, these findings confirm that combining LSTM’s sequential modeling capabilities with a multi-task learning framework significantly boosts the model’s effectiveness in providing accurate and personalized learning path recommendations.

\subsubsection{Impact of learning path length on performance}

Figure~\ref{fig:evaluation-metrics} demonstrates how the length of the recommended learning path affects model performance across Accuracy, F1 Score, Precision, and Recall. As expected, performance generally decreases for all models as the learning path length increases from 3 to 9, highlighting the increased complexity and uncertainty that comes with predicting longer sequences of learning activities.

Comparing architectures, LSTM-based models such as LSTM and Seq2Seq + LSTM consistently outperform their Standard RNN counterparts across all metrics and path lengths. The addition of Seq2Seq and attention mechanisms brings modest improvements, particularly when used with LSTM, indicating their benefit in modeling complex dependencies. In contrast, models relying on the Standard RNN backbone, especially Seq2Seq + Standard RNN and Seq2Seq + Standard RNN + Attention, show more significant performance drops as the path length increases, revealing their weaker capacity for handling longer-term dependencies in sequence prediction.

Importantly, the proposed method demonstrates strong and stable performance that often rivals or exceeds several single-task baselines, especially for shorter paths. When viewed as a form of multi-task learning, where the model simultaneously learns to recommend appropriate next concepts and predict learning performance, it highlights how adding an auxiliary constraint can guide the model toward generating higher-quality learning paths. This multi-task perspective shows that integrating domain-specific constraints into the learning objective can be a practical and effective way to improve recommendation quality, complementing architectural advances like LSTM and attention for the learning path recommendation.

\section{Related Work}
\label{sec5}
Learning path recommendation has been advanced with deep learning. For instance, Yun \textit{et al.} propose a novel method of offline reinforcement learning called Doubly Constrained deep Q-learning Network (DCQN)~\cite{yun2024doubly}. This method utilizes two generative models to fit existing student historical interaction data, which in turn, constrains the original policy network to generate new actions based on past interactions, avoiding the occurrence of overestimated actions and reducing extrapolation errors. Shou \textit{et al.} propose a learning path planning algorithm that uses collaborative analysis of online learning behavior to build a concept interaction model, a directed learning path network, and measures local structure similarity between knowledge nodes~\cite{shou2020learning}. By calculating learning behavior similarity using a Kullback-Leibler divergence matrix and clustering learners with similar behaviors, the algorithm generates personalized optimal learning paths, which are empirically validated with online behavior and test data. Meng \textit{et al.}  introduces a new learning path generation method called learning diagnosis–learning path (LD–LP)~\cite{meng2021ld}, which adapts to students by considering knowledge structure, learner ability, learning time, and repetition. Unlike previous approaches, it adjusts knowledge difficulty based on students’ performance and time spent, generating personalized learning sequences, and was tested with mathematics majors.

Recent years, large language models (LLMs) has emerged to further enhance the learning path recommendation. For example, Ng \textit{et al.}  presents a new approach that uses LLMs combined with prompt engineering to generate personalized and pedagogically sound learning paths by embedding learner-specific information into the prompts~\cite{ng2024educational}. Experimental results show that this method, especially with GPT-4, significantly outperforms a baseline in accuracy, user satisfaction, and learning path quality, with long-term analysis confirming its potential to enhance learner performance and retention. Li \textit{et al.}  introduces SKarREC, a Structure and Knowledge Aware Representation learning framework for concept recommendation that combines factual knowledge from LLMs with concept relationships from a knowledge graph to build richer textual representations~\cite{li2024learning}. It uses a graph-based adapter, pre-trained with contrastive learning and fine-tuned for recommendation, creating a text-to-knowledge-to-recommendation pipeline that outperforms previous adapters in concept recommendation tasks.

However, few previous studies have fully investigated multi-task learning for learning path recommendation, which aims to enhance performance by leveraging benefits from related tasks. This study addresses this gap by proposing a multi-task learning framework that jointly optimizes learning path recommendation alongside the deep knowledge tracing  to improve overall recommendation accuracy and robustness.

\section{Conclusion and Future Work}
\label{sec6}

This study introduces a multi-task LSTM model designed to improve learning path recommendation by utilizing shared information across related tasks. It reformulates learning path recommendation as a Seq2Seq prediction task that generates personalized paths based on a learner’s historical data. The model incorporates a shared LSTM layer to extract common features for both learning path recommendation and deep knowledge tracing, supplemented by task-specific LSTM layers to address each objective individually. Additionally, a non-repeat loss is applied to discourage duplicate items within the generated learning path. Experimental results show that the proposed method outperforms all baselines across all evaluation metrics, achieving an Accuracy of 0.3489, F1 Score of 0.3241, Precision of 0.3202, and Recall of 0.3489. This consistent improvement demonstrates that the approach brings substantial enhancements, enabling more effective modeling of student learning paths than traditional RNN, LSTM, and Seq2Seq-based models. In the future, we plan to enhance the proposed method by incorporating knowledge extracted from Large Language Models (LLMs), which can provide richer contextual information and domain-specific insights to further refine learning path recommendations and better adapt to diverse learner needs.
%

\section*{Acknowledgment}

This research work is supported by NSF  under award numbers 2235731 and 2401860,  and by the Army Research Office under Cooperative Agreement Number W911NF-24-2-0133. The views and conclusions contained in this document are those of the authors and should not be interpreted as representing the official policies, either expressed or implied, of the NSF or the Army Research Office or the U.S. Government. The U.S. Government is authorized to reproduce and distribute reprints for Government purposes notwithstanding any copyright notation herein. Additionally, they acknowledge the use of AI-based tools, such as ChatGPT, for assistance in editing, grammar enhancement, and spelling checks during the preparation of this manuscript.



\bibliographystyle{IEEEtran}
\bibliography{References}
%



\end{document}